\documentclass[%
 amsmath,amssymb,
 aps,
 prfluids,
]{revtex4-2}

\usepackage{tikz}
\usepackage{lineno,hyperref}
\usepackage{epsfig}
\usepackage{amssymb}
\usepackage{mathtools}
\usepackage{booktabs} 
\usepackage{enumitem}
\usepackage{graphicx}
\usepackage{subcaption}
\usepackage[ruled,vlined]{algorithm2e}
\usepackage{pgfplots}
\modulolinenumbers[5]
\usepackage{dcolumn}
\usepackage{comment}
\usepackage[percent]{overpic}
\usepackage{bm}

\def\mica#1{{\textcolor{black}{#1}}}  
\begin{document}


\title{Clustering of vector nulls in homogeneous
isotropic turbulence}

\author{D.O.~Mora$^{1,2}$, M.~Bourgoin$^3$, P.D.~Mininni$^4$ and M.~Obligado$^1$}
\affiliation{$^1$ Universit\'{e} Grenoble Alpes, CNRS, Grenoble-INP, LEGI, F-38000, Grenoble, France}
\affiliation{$^2$ Department of Mechanical Engineering, University of Washington, Seattle, Washington 98195-2600, USA}
\affiliation{$^3$ Univ Lyon, Ens de Lyon, Univ Claude Bernard, CNRS, Laboratoire de Physique, 46 all\'ee d’Italie F-69342 Lyon, France}
\affiliation{$^4$Universidad de Buenos Aires, Facultad de Ciencias Exactas y Naturales, Departamento de F\'\i sica, \& IFIBA, CONICET, Ciudad Universitaria, Buenos Aires 1428, Argentina.}

\begin{abstract}
  We analyze the vector nulls of velocity, Lagrangian acceleration, and vorticity, coming from direct numerical simulations of forced homogeneous isotropic turbulence at $Re_\lambda \in O([40-600])$.  We show that the clustering of velocity nulls is much stronger than those of acceleration and vorticity nulls. These acceleration and vorticity nulls, however, are denser than the velocity nulls. We study the scaling of clusters of these null points with $Re_\lambda$ and with characteristic turbulence lengthscales. We also analyze datasets of point inertial particles with Stokes numbers $St = 0.5$, 3, and 6, at $Re_\lambda = 240$. Inertial particles display preferential concentration with a degree of clustering that resembles some properties of the clustering of the Lagrangian acceleration nulls, in agreement with the proposed sweep-stick mechanism of clustering formation.
\end{abstract}


\maketitle 

\section{Introduction}

Single-phase and particle-laden turbulent flows are of interest in many industrial, natural and environmental situations. But despite their relevance, there are still many open questions that severely limit our understanding of these flows. For instance, the study of geometrical properties of the velocity, the Lagrangian acceleration, and the vorticity fields in turbulent flows has received considerable attention in the last decades. The geometrical properties of these fields can be useful to model important phenomena in turbulent flows such as superdifusivity, preferential concentration of particles,  vortex reconnection, among many others. Some works, focused on characterizing simply connected regions of vorticity, have found that such regions tend to cluster \cite{ishihara2013thin, moisy2004geometry, tanahashi2008fine, itoh2018large}. Other studies have focused on the vector field \textit{nulls} \cite{goto2004particle, davila2003richardson, mcgavin2019reconnection}, the points where the modulus of vectorial quantities is equal to zero (i.e., the set of points $\mathbf{X}_p =\{ \mathbf{x_n}=(x_n,y_n,z_n) \in \Re^3 \, | \, \mathbf{p}(x_n,y_n,z_n)=0\}$, where $\mathbf{p}$ is some vector field as, e.g., the vorticity). It is nevertheless unclear how these quantities relate to each other and, furthermore, their dependence on different parameters such as the Reynolds number or the homogeneity and isotropy of the underlying flow.

In spite of the limited knowledge on their spatial distribution and scaling properties, the geometry of these vector nulls, also known in some cases as stationary or fixed points, has successfully been related to the underlying physics of turbulence and to turbulence-particle interactions. For instance, their statistics and scaling properties have been associated with the fractal nature of turbulence, and scaling laws for the distribution of vector nulls have been derived from fractal dimensions of the velocity field compatible with Kolmogorov scalings \cite{davila2003richardson, goto2004particle, chen2006turbulent}. Also, it has been noted that the velocity nulls carry information about the turbulent kinetic energy dissipation rate $\varepsilon$ (see \cite{sreenivasan1983zero,goto2009dissipation,liepmann1953counting, davila2003richardson, goto2009dissipation}) and can be used to model relative pair dispersion~\cite{faber2009_PoF}. Furthermore, the Lagrangian acceleration nulls have been related to the degree of preferential concentration found when inertial particles are added to a turbulent flow \cite{coleman2009unified,goto2004particle,uhlmann2017clustering}.

Some specific results for three choices of these null points are worth mentioning. On the one hand, it has been shown that properties of the set of velocity nulls (satisfying $\mathbf{v}(\mathbf{x_n})=0$ where $\mathbf{v}$ is the fluid velocity field, and referred in the following as ``stagnation points" or STPS) from a one-dimensional (1D) measurement, can be related to the Taylor microscale  $\lambda$ of homogeneous isotropic turbulence (HIT) via the Rice theorem \cite{rice1945mathematical,liepmann1953counting}. More recently, Goto and Vassilicos extended these results to three-dimensional (3D) fields \cite{goto2009dissipation}, \textcolor{black}{ i.e., the showed that the average distance beteen STPS in 3D is also proportional to the Taylor length-scale}. On the other hand, the set of Lagrangian acceleration nulls (with $\mathbf{a}(\mathbf{x_n})=0$, and referred in the following as ZAPS for ``zero acceleration points"), are at the core of the proposed \textit{sweep-stick mechanism} \cite{coleman2009unified}, that models preferential concentration in inertial-particle-laden flows. This model suggests that the inertial particles mimic the spatial distribution of ZAPS, for particles with a Stokes number $St$ above unity \cite{Obligado2014, sumbekova2017preferential, uhlmann2017clustering}. Finally, the vorticity nulls ($\pmb{\omega}(\mathbf{x_n})=0$, or WZERO in the following), have been related to vortex reconnection events and the turbulent cascading process \cite{yao2020physical, mcgavin2019reconnection}. Furthermore, low vorticity regions are also expected to control the centrifugal expulsion of inertial particles (and their clustering) for $St<1$. Within this mechanism, dense particles are expected to be expelled from the core of eddies and to accumulate in regions of high strain and low vorticity \cite{wang1993settling}.

Despite the interest and ongoing research in the topic, to the authors best knowledge no joint systematic study on the global properties and $Re_\lambda$-dependence of the clustering of these points has been carried out (where $Re_\lambda$ is the Reynolds number based on the flow Taylor microscale), nor of their relation with inertial particle concentration fields. Vassilicos and collaborators have conducted the most extensive studies on the geometry of STPS and ZAPS using pair distribution functions \cite{faber2010acceleration, chen2006turbulent, coleman2009unified,davila2003richardson, goto2006self}.  These works have provided analytical predictions on these points' statistical properties, and some recent studies \cite{Obligado2014, uhlmann2017clustering, baker2017coherent} have retrieved some evidence that partially validates some of these predictions. Considering that Vassilicos and collaborators used pair distribution functions to examine the vector nulls clustering, they could not examine the local vicinity around a null point so that conditioned statistics could be computed and related to surrounding turbulent phenomena (e.g., to preferential concentration). In this respect, the study of Obligado and collaborators~\cite{Obligado2014} found that the 2D spatial distribution of inertial particles with $St=2$ and $4$ (characterised via Vorono\"i tessellations) presents similarities with the regions of low Lagrangian acceleration.

In this work we analyze velocity, Lagrangian acceleration, and vorticity nulls through Vorono\"{i} tessellations \cite{ferenc2007size}. The fields examined come from forced direct numerical simulations (DNSs) of HIT. We study seven different DNS datasets, exploring a wide range of Reynolds numbers ($Re_\lambda \in [40-610]$) and different forcing schemes. Vorono\"{i} tessellations, contrary to pair correlation functions, allow the examination of the nulls local ``concentration" maps and cluster size distributions, similar to previous studies of clustering of inertial particles \cite{Monchaux2010, falkinhoff2020preferential}. Our results show that the degree of clustering of STPS is much larger than the respective ones for ZAPS or WZERO (while the concentration follows an inverse trend), and display a clear scaling with $Re_\lambda$. The results also confirm that the concentration of STPS is at least one order of magnitude smaller than those of ZAPS or WZERO, in agreement with the scalings of Chen et al.~\cite{chen2006turbulent}.  Also, while the properties of ZAPS and WZERO have a similar trend with $Re_\lambda$, STPS spatial structure presents a different dependence with this parameter.

Finally, for one of our DNS (with $Re_\lambda=240$) we also studied the behaviour of dense, point-like inertial particles. The objective is to generalise the results from the previous work of Obligado and collaborators \cite{Obligado2014}. In this aspect, the novelty of the present work compared to previous ones is twofold: we generalise the study to 3D Vorono\"i tessellations (thus eliminating any bias caused by projecting fields to 2D) and instead of studying regions with low vorticity or acceleration, we use an interpolation method that allows to actually detect nulls as points (as detailed in Sec.~\ref{sc:nullsc}). We then compare the clustering properties of different nulls datasets with the clustering of inertial particles with Stokes numbers $St$ of $0.5$, 3 and 6. We find that, in agreement with previous works, the spatial segregation of particles with $St>1$ presents larger similarities with the ZAPS than with the other vector nulls, consistently with the behavior expected from the sweep-stick mechanism. 

\section{Methodology}

\subsection{Numerical simulations \label{sec:numerics}}

Our numerical datasets of the Eulerian velocity, Lagrangian acceleration, and Eulerian vorticity came from DNSs. These simulations follow standard practices regarding their temporal integration, de-aliasing procedures, and have an adequate spatial resolution of the smallest scales, i.e., $\kappa\eta \gtrsim 1$ \cite{pope_2000}. Here $\eta$ is the Kolmogorov lengthscale, $\eta = (\nu^3/\varepsilon)^{1/4}$ (where $\varepsilon$ is the kinetic energy dissipation rate, and $\nu$ the kinematic viscosity of the fluid), and $\kappa=N/3$ the maximum resolved wavenumber in Fourier space (with $N$ the linear spatial resolution).  Fully dealiased pseudospectral methods with second-order Runge-Kutta methods for the time stepping are used. The 3D simulation domain for all datasets has dimensions of $2\pi\times2\pi\times2\pi$. All relevant simulation parameters can be found in Table \ref{tab:dnspara}.

Numerical simulations solve the incompressible Navier-Stokes equations for the velocity ${\bf v}$ with a random solenoidal forcing ${\bf f}$,
\begin{equation}
    \frac{D {\bf v}}{Dt} = \frac{\partial {\bf v}}{\partial t} + {\bf v} \cdot \pmb{\nabla} {\bf v} = - \pmb{\nabla} p' + \nu \nabla^2 {\bf v} + {\bf f},
    \label{eq:NS}
\end{equation}
where $p' = p/\rho$ (with $p$ is the pressure and $\rho$ a uniform mass density), which is obtained from the incompressibility condition $\pmb{\nabla} \cdot {\bf v}=0$. In Eq.~(\ref{eq:NS}), $D {\bf v}/Dt = {\bf a}$ is the Lagrangian acceleration of the fluid elements, while the vorticity field is given by $\pmb{\omega} = \pmb{\nabla} \times {\bf v}$. We define the r.m.s.~velocity as $u'=\left<|v_i|^2\right>^{1/2}$ (where $v_i$ is a Cartesian component of the velocity), the Taylor scale as $\lambda=(15 \nu u'^2/\varepsilon)^{1/2}$, and the integral scale as ${\cal L} = \pi/(2 u'^2) \int{ E(k)/k \, dk}$ (where $E(k)$ is the isotropic energy spectrum).

\begin{table}[b]
\begin{ruledtabular}
\begin{tabular}{l c c c c c}
\hfill
Dataset  &  $N$  &$\mathcal{L}/(2\pi)$ &$10^{3}\eta$ & $Re_\lambda$ & $\#$ snapshots \\
\hline
DNS-64   &     64 &               0.304 &                50 &           40 &             80 \\
DNS-128  &    128 &               0.291 &                24 &           70 &             50 \\
DNS-256  &    256 &               0.291 &                12 &          120 &             50 \\
DNS-512  &    512 &               0.238 &                 6 &          240 &             20 \\
DNS-1024 &   1024 &               0.309 &                 3 &          520 &              9 \\
JHU-1024 &   1024 &               0.217 &             $2.8$ &          430 &             15 \\
JHU-4096 &   4096 &               0.221 &             $1.4$ &          610 &              1 \\
\end{tabular}
\end{ruledtabular}
	\caption{DNS parameters. $N$ is the number of points in each direction, such that $N^3$ is the total number of grid points in the simulation domain. $\mathcal{L}/(2\pi)$ is the integral lengthscale in units of the domain linear size of length $2\pi$. $\eta$ is the Kolmogorov dissipation scale, multiplied by $10^3$ for convenience. $Re_\lambda$ is the Reynolds number based on the Taylor microscale $\lambda$, and $\#$ snapshots is the number of snapshots of the vector fields used for the analysis. \label{tab:dnspara}}
\end{table}

We use a total of seven numerical datasets. The first five datasets (labeled in the following as ``DNS-$N$", where $N$ is the linear resolution of each dataset) were obtained using the \textsc{Ghost} code (see \cite{mininni2011parallel, Rosenberg_2020} for further details of the code). In this case, the solenoidal forcing ${\bf f}$ is given by a superposition of Fourier modes with random phases in the shell with wavenumber $k=1$. A new random forcing was generated every $0.5$ large-scale turnover times, and the forcing was linearly evolved from its previous state to the next state along this period of time. This results in a continuous and slowly evolving random forcing with correlation time of $0.5$ turnover times, which at the largest resolution considered has an integral scale $\mathcal{L}/(2\pi) \approx 0.309$, and which will be useful for simulations with inertial particles as discussed below. These simulations also use the largest Reynolds number attainable at their given spatial resolution, with $\kappa\eta\approx 1$ (see Table \ref{tab:dnspara}). The last two datasets (labeled ``JHU-$N$") correspond to simulations of homogeneous and isotropic turbulence from the Johns Hopkins Turbulence (JHU) database \cite{li2008public}, at $Re_\lambda$ similar to the largest simulation in the ``DNS" dataset. \textcolor{black}{It is worth noting that the JHU database does not provide the Lagrangian acceleration or the vorticity field as primary globally accessible variables; these fields can be queried and computed instead at each point in the physical space via finite differences \cite{li2008public}. Due to limited computational resources, and to the different numerical errors between finite differences and pseudospectral computation (used for all the other simulations and fields in this study), we only computed for the JHU database the STPS from the available velocity fields.} In these simulations, the forcing ${\bf f}$ keeps the kinetic energy constant in Fourier shells with $k \leq 2$. For dataset JHU-4096 \cite{Yeung_2012}, this results in an integral scale $\mathcal{L}/(2\pi) \approx 0.221$. The last simulation is better resolved, with a value of $\kappa\eta$ approximately $2.7$ times larger than the ``DNS-$N$" datasets. While resolution has been found to impact significantly in some statistical properties of turbulence, as, e.g., the scaling of extreme field gradients \cite{Donzis_2010, Buaria_2019}, we will see that it does not seem to affect so significantly the statistics of clustering of nulls of the vector fields.

In particular, the numerical simulations DNS-1024, JHU-1024, and JHU-4096 have similar values of $Re_\lambda$ (between $430$ and $610$), but different values of $\kappa\eta$ (with, as already mentioned, the JHU simulations being better resolved). These simulations can thus be used to partially consider the effect of resolution in the statistics of null points. As will be shown below, we obtain consistent results in all simulations, but with the simulations with larger values of $\kappa\eta$ having a slightly larger number of null points. Note that while {\it a priori} it could be expected that less resolved simulations should have more vector field nulls (as a result of Gibbs phenomena if the vector fields are not well resolved), the simulations display the opposite behavior. This indicates that Gibbs phenomena is not dominant even in the simulations with $\kappa\eta \approx 1$. Instead, the increase in the number of null points with larger values of $\kappa\eta$ indicate that as the vector fields become more intermittent, the number of zeros in the fields increases. Previous studies comparing simulations with different values of $\kappa\eta$ in a different context \cite{Wan_2010, Donzis_2010} also indicate that this can be the case: the simplest criterion that the Kolmogorov scale should be resolved yields simulations with accurate estimations of the statistical lower-order moments of the vector fields, but much more stringent conditions are needed to capture higher-order statistical moments, with intermittency increasing as $\kappa\eta$ increases \cite{Wan_2010}. \textcolor{black}{However, the DNS and JHU simulations also have different forcing schemes, which are also known to affect the distribution and number of null points \cite{weiss2019impact}. Thus, a detailed study of the effect of varying the forcing mechanism, and of the effect of varying $\kappa\eta$ for a fixed forcing scheme on the statistics of nulls, would also be of interest. The fractal properties of many field nulls were considered in experimental data \cite{sreenivasan1983zero, kailasnath1993zero}, and in numerical data for different flows \cite{Ott_1992, Sorriso_2002, Rodriguez_Imazio_2010} but in the simulations mostly for cases with $\kappa\eta \approx 1$. We thus warn the reader of the limitations in our knowledge of the effect of both varying the forcing and the spatial resolution, and leave a detailed study of these effects for a future work.}

For DNS-512 we also have data of $10^6$ inertial point particles without gravity, which will be considered in Sec.~\ref{sec:part}. Particles are integrated following the equations
\begin{equation}
    \frac{d{\bf x}_p}{dt} = {\bf v}_p, \;\;\;\;\;\; \frac{d{\bf v}_p}{dt} = \frac{1}{\tau_p}\left[ {\bf v}({\bf x}_p) - {\bf v}_p \right],
\end{equation}
where ${\bf x}_p$ is the particle position, ${\bf v}_p$ the particle velocity, ${\bf v}({\bf x}_p)$ the fluid velocity at the particle position, and $\tau_p$ the Stokes time. The Stokes number of the particles is then defined as $St = \tau_p/\tau_\eta$, where $\tau_\eta$ is the eddy turnover time at the Kolmogorov scale. These equations are integrated with a high-order Runge-Kutta method to evolve in time, and a high-order three-dimensional spatial spline interpolation to estimate the fluid velocity ${\bf v}({\bf x}_p)$ at the particle position (see \cite{Yeung_1988, Angriman_2020} for details).

The Taylor-based Reynolds number, $Re_\lambda=u' \lambda/\nu$, spans one and a half decades. We have $Re_\lambda \in [40,610]$ for spatial resolutions of 64$^3$, 128$^3$, 256$^3$, 512$^3$, 1024$^3$, and 4096$^3$ grids points. We took enough snapshots of the vector fields to have adequate global statistics. The JHU datasets were post-processed using the Sciserver platform \cite{taghizadeh2020sciserver}, and we used the Python library \textsc{Freud} \cite{freud} to compute the 3D Vorono\"{i} diagrams for all datasets.
 
\subsection{Nulls calculation}
\label{sc:nullsc}
We applied the method proposed by Haynes and collaborators \cite{haynes2007trilinear,haynes2010method,murphy2015appearance} to compute our data vector nulls. Although they developed this method for magnetic fields, recent studies have used the same method to compute nulls of vorticity fields \cite{mcgavin2019reconnection}. We briefly describe this algorithm main steps: First, for each cell in the domain, we survey its vector values at the cell's corners (i.e., at the grid points) to determine if there is a change of sign in all components of the target vector field (i.e., we survey the 8 corners of the target cubic cell). For a properly resolved and dealiased DNS, if any of the $x$, $y$, or $z$ components of the field do not change in sign within the cell, there cannot be a zero inside it. If, on the other hand, there are changes of sign in every vector component inside the cell (i.e., there is a change of sign in any of the 8 corners of the cubic cell, and for each vector component  $x$, $y$, and $z$), we use the 8 corner component values to feed a trilinear interpolation algorithm, and thereby, we build a local vector interpolation function. Then, we proceed by feeding this function into a Newton-Raphson method \cite{press1989numerical} to verify if there is a zero within the cell. This algorithm is somewhat similar to those proposed by Vassilicos and collaborators \cite{dallas2009stagnation, coleman2009unified}, and can be easily parallelized.

However, considering the non-linearity and resulting spatial complexity of the turbulent fields here studied, there is a caveat: some of the cells' zeros can be located outside the target cell. Haynes and Parnell \cite{haynes2007trilinear} propose that these zeros could be accepted if the zeros' locations are not very far from the \textit {local cell}.
Although these ``satellite" zeros increase the nulls density (and improve the statistics), we opted for a more conservative approach and considered as valid nulls only those zeros found inside the target cell. We took this decision based on a benchmark calculation that showed that including these ``satellite" nulls may lead to pathological behaviors of the global parameters coming from the Vorono\"{i} tessellation analyses.

\section{Nulls analysis}
\label{sc:nall}
We computed the nulls of velocity, Lagrangian acceleration, and vorticity following the Haynes algorithm for all the data in table \ref{tab:dnspara}. \textcolor{black}{Table \ref{tab:cells} summarizes the total number of zeros found by this algorithm (for all snapshots in a given simulation), as well as the average number of zeros per field snapshot in each simulation.} We then applied the 3D Vorono\"{i} tessellation analysis on these nulls positions. This analysis followed the same protocol of studies focusing on inertial particle clustering \cite{Monchaux2010,Obligado2014}. Thus, we quantified their degree of clustering via the standard deviation $\sigma_\mathcal{V}$ of the normalized Vorono\"{i} cells volume $\mathcal{V}=V/\langle V\rangle$ (where $\langle V\rangle$ denotes the average volume), which ultimately quantifies the effects of the 
``voids" (i.e., of low density regions  \cite{sumbekova2017preferential}) present in the nulls spatial distribution. We consider that clustering is present when $\sigma_\mathcal{V}>\sigma_\textrm{RPP}$ \cite{Monchaux2010}, where $\sigma_\textrm{RPP}\approx 0.42$ is the standard deviation of a 3D random Poisson process (RPP), which has no correlations at any scale \cite{tanemura2003statistical}.

\textcolor{black}{
\begin{table}[b]
\begin{ruledtabular}
\begin{tabular}{l|rc|rc|rc}
\hfill
Dataset & total $\#$ of STPS & $\langle \mathrm{STPS}\rangle_{snap}$ & total $\#$ of ZAPS & $\langle \mathrm{ZAPS}\rangle_{snap}$ & total $\#$ of WZERO & $\langle \mathrm{WZERO}\rangle_{snap}$ \\
\hline
DNS-64   &                        1528 &      64 $\pm$ 27 &                       35956 &     443 $\pm$ 163 &                        141026 &     1720 $\pm$ 520 \\
DNS-128  &                        12431 &     264 $\pm$ 100 &                      193742 &    2849 $\pm$ 1000 &                       976850 &    14365 $\pm$ 4900 \\
DNS-256  &                       34000 &     700 $\pm$ 300 &                       886963 &    17931 $\pm$ 6500 &                      5108326 &    102166 $\pm$ 33858\\
DNS-512  &                       85598 &    5708 $\pm$ 3700 &                     5155101 &  343673 $\pm$ 165719 &                     26868036 &  2239003 $\pm$ 776174 \\
DNS-1024 &                       65537 &    7282 $\pm$ 1188 &                     6040950 &  1006825 $\pm$ 30143  &                     38000000 &   6024043 $\pm$ 1265305\\
JHU-1024 &                      258849 &   19911 $\pm$ 500 &                           -- &       -- &                            -- &        -- \\
JHU-4096 &                      100000 &  100000 &                           -- &       -- &                            -- &       -- \\
\end{tabular}
\end{ruledtabular}
	\caption{\textcolor{black}{Number of zero nulls found for each vector field (STPS, ZAPS and WZERO) and simulation. Both total numbers of zeros (i.e., aggregated for all times available) and averages per snapshot with their respective standard deviation (indicated by the $\pm$ deviation) are given. The averages just mentioned, indicated by angular brackets, refer to the average number of the respective zeros per field snapshot.}\label{tab:cells}}
\end{table}}

Analogous to inertial particle studies, we also computed the clusters volume probability distribution function (PDF) via the algorithm proposed by Monchaux et al.~\cite{Monchaux2010}. We selected from the volume cells collection those cells that are below a threshold $\mathcal{V}_\textrm{th}$, and considered as clusters the groups of two or more of those cells sharing a boundary (face). We picked $\mathcal{V}_\textrm{th}$ as the location of the first crossing (i.e., for $\mathcal{V}<1$) between an RPP PDF and the Vorono\"{i} cells' PDF. In our analysis we took $\mathcal{V}_\textrm{th} \approx 0.5$ (and $\mathcal{V}_\textrm{th} \approx 0.56$ for the inertial particles in Sec.~\ref{sc:nullsc}). Interestingly, this threshold did not depend strongly on $Re_\lambda$ or on the dataset we analyzed. We note, however, that there were not sufficient clusters in single snapshots of the DNS-64 dataset to reach adequate statistics.

Nevertheless, there is a discrepancy in the values of $\mathcal{V}_\textrm{th}$ between our simulations and the JHU datasets, as will be shown later. \textcolor{black}{This is partially related to the fact that the JHU datasets have a larger number of nulls. Although it could be expected both JHU and our DNS datasets should strictly follow the same power-law fitting exponents, it is worth pointing out that due to the their different large-scale forcing methods, this may not the case. This argument is supported by the simulations and theoretical predictions of Goto and Vassilicos \cite{goto2009dissipation} that show that the stagnation point structure changes depending on the forcing method, and the forcing wave number. In other words, even simulations with similar $Re_\lambda$ but different forcing may exhibit different topologies (besides the effect of spatial resolution already discussed in Sec.~\ref{sec:numerics})}. These authors further argued that differences in STPS topology may impact the turbulent cascading process via changes in the normalized dissipation rate $C_\varepsilon$. Weiss et al.~\cite{weiss2019impact} also reported that the degree of particle clustering depends on the forcing used to sustain the turbulence.  Taking into account that Coleman and Vassilicos have linked particle clustering to properties of the ZAPS, the effect of spatial resolution, and the results of Weiss and collaborators that highlight the non-negligible influence of the large scales on the turbulent field topology, we consider that such a discrepancy is to be expected. Moreover, as will be shown next, once $\mathcal{V}_\textrm{th}$ is defined as described above, other results from all the datasets are compatible between themselves.

\subsection{Scaling of averaged quantities for all vector field nulls}

\begin{figure}
    \begin{center}

		\begin{subfigure}[t]{0.48\textwidth}
			\begin{overpic}[scale=0.55, trim=0cm 1.5cm 0cm 0cm]{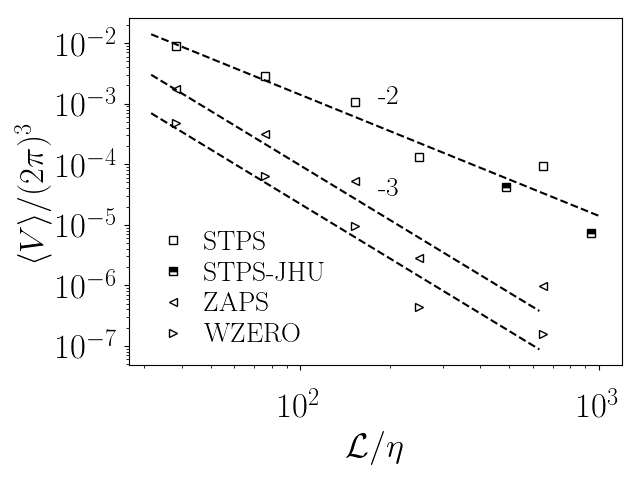}
            \put (89,55) {\huge a)}
            \end{overpic}
		 	\caption{\label{fig:MEANVORO-F}}
		\end{subfigure}
						~
		\begin{subfigure}[t]{0.48\textwidth}
			\begin{overpic}[scale=0.55, trim=0cm 1.5cm 0cm 0cm]{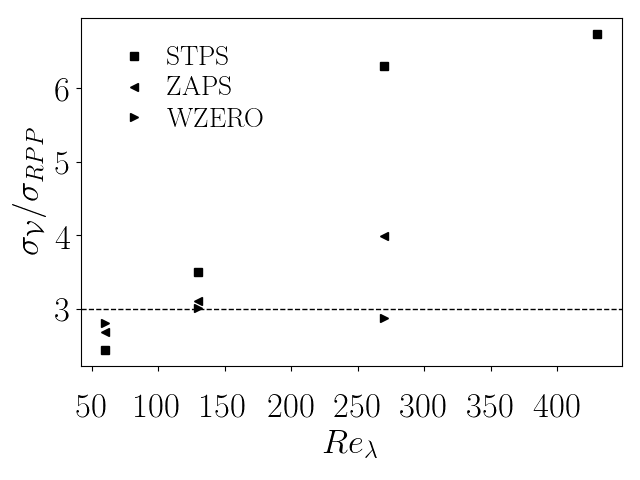}
            \put (89,15) {\huge b)}
            \end{overpic}
			\caption{\label{fig:STD-F}}
		\end{subfigure}

		\begin{subfigure}[t]{0.48\textwidth}
			\begin{overpic}[scale=0.55, trim=0cm 1.5cm 0cm 0cm]{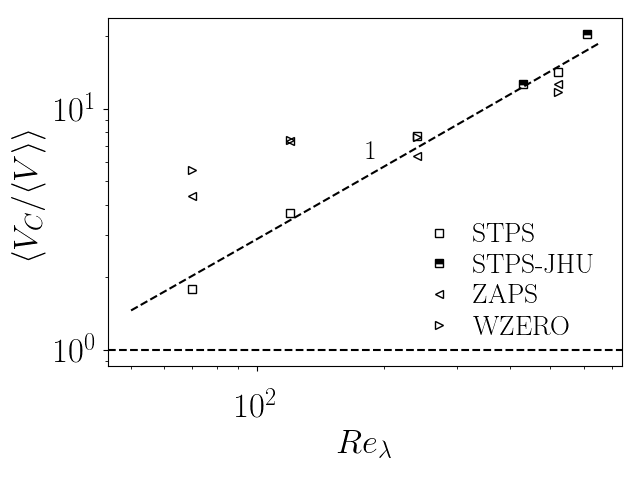}
             \put (20,55) {\huge c)}
            \end{overpic}
		    \caption{\label{fig:CLUM-MEAN}}
		\end{subfigure}
						~
		\begin{subfigure}[t]{0.48\textwidth}
			\begin{overpic}[scale=0.55, trim=0cm 1.5cm 0cm 0cm]{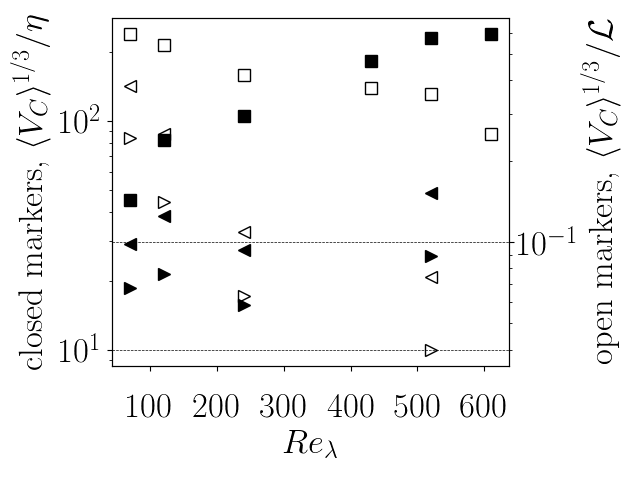}
            \put (72,15) {\huge d)}
            \end{overpic}
		    \caption{\label{fig:CLUM-LINT}}
		\end{subfigure}
		
	\end{center}
	\caption{Global Vorono\"{i} statistics for the different field nulls. a) Average Vorono\"{i} volume sizes of STPS, ZAPS, and WZERO. STPS for the JHU datasets (``STPS-JHU") are indicated by a different marker here an in the following panels. b) Standard deviation of the Vorono\"{i} cell volumes with respect to the one from a Poisson distribution (RPP). c) Average cluster volume normalized by the average cell volume. d) Average cluster size over Kolmogorov ($\eta$, left vertical axis with closed markers) and integral length scales ($\mathcal{L}$, right vertical axis with open markers). Markers shapes are the same for all panels. Power laws and some reference values are indicated by straight lines.}
\end{figure}

Vassilicos and collaborators \cite{davila2003richardson, goto2004particle, chen2006turbulent} report that the number density ($n_s\sim\langle V\rangle^{-1}$, i.e., the inverse of the average Vorono\"{i} cell volume) of 3D STPS and of 3D ZAPS scale as $n_s\sim\big(\mathcal{L}/\eta\big)^\delta$, where $\delta$ (a fractal dimension) takes the values of 2 and 3 respectively for each set of nulls. For STPS, this fractal dimension can be seen as a consequence of viewing turbulence as a self-similar process. Under such assumption, the energy spectrum exponent (i.e., the -5/3 power law) can be related to the fractal exponent $\delta$ via Orey's theorem leading to $\delta=2$ for the 3D STPS (see \cite{davila2003richardson,chen2006turbulent}).   Likewise, Moisy and Jimenez \cite{moisy2004geometry} report a box dimension for the number density of vortical structures (resp.~WZERO points) close to $-3$. Our results for the average Vorono\"{i} volume size ($\langle V\rangle\sim n_s^{-1}$) of the  different null points are consistent with the mentioned scalings and observations (see Fig.~\ref{fig:MEANVORO-F}). However, we may not have enough scale separation in the inertial range (only a decade in terms of $\mathcal{L}/\eta$) to ascertain without doubt their exact numerical values. Moreover, our results do  reveal that the STPS are indeed very scarce (\textcolor{black}{i.e., they have a smaller concentration in space}) when compared to ZAPS or WZERO, and thereby, their larger average Vorono\"{i} cell volume, which as already mentioned is inversely proportional to the number density of the respective nulls: In other words, note from Fig.~\ref{fig:MEANVORO-F} that $\langle V \rangle\vert_\mathbf{v=0}>\langle V \rangle\vert_\mathbf{a=0}>\langle V \rangle\vert_\mathbf{\omega=0}$ for all $Re_\lambda$ considered.

\textcolor{black}{The Vorono\"{i} volume standard deviation for the nulls, which quantifies the degree of clustering via Vorono\"{i} tessellations, except the for smallest value of $Re_\lambda$, roughly satisfies a similar ordering as the mean (see Fig.~\ref{fig:STD-F}), i.e., ${\sigma_\mathcal{V}}\vert_\mathbf{v=0} > {\sigma_\mathcal{V}}\vert_\mathbf{a=0} > {\sigma_\mathcal{V}}\vert_\mathbf{\omega=0}$. In other words, the stagnation points become more ``clustered" than the acceleration or vorticity nulls}. This is consistent with the findings of Chen et al.~\cite{chen2006turbulent}, who used pair distribution functions to characterize the geometry of STPS and ZAPS. At increasing $Re_\lambda$, the velocity nulls (STPS) cluster more strongly than the acceleration and vorticity nulls, with ${\sigma_\mathcal{V}}\vert_\mathbf{v=0}$ growing with $Re_\lambda$ (a power law is indicated in the figure as a reference). The increased complexity (at all scales) of the STPS topological structure is reflected in its respective PDF (see Fig. \ref{fig:VOROPDF-STP} and the discussion in Sec.~\ref{sc:pdf}), which shows that at increasing $Re_\lambda$ a power law close to -5/3 emerges in the PDF of the Vorono\"i volumes of these nulls. In addition, the standard deviations of the cell volumes of vorticity and acceleration nulls depend weakly on $Re_\lambda$, if at all. \textcolor{black}{As will be shown later, this is a consequence of the behavior of their respective PDFs (see Figs.~\ref{fig:VOROPDF-ZAPS} and \ref{fig:VOROPDF-WZERO}). These PDFs, for different values of $Re_\lambda$, roughly collapse for $V/\langle V\rangle>1$ (i.e., for cells with volumes larger than the average), which correspond to the cells that contribute the most to $\sigma_\mathcal{V}$. }

Interestingly, at our smallest values of $Re_\lambda$, the ZAPS and WZERO exhibit similar numerical magnitudes of the standard deviation and of the normalized average cluster size $\langle V_C \rangle /\langle V \rangle$ (see Fig.~\ref{fig:CLUM-MEAN}, where $V_C$ denotes the volume of the clusters). In other words, under the Vorono\"{i} analysis criteria, both fields display a similar degree of clustering. \mica{We note however that this statistical signature does not necessarily imply that both fields coincide.} \textcolor{black}{This observation is indeed contentious, as in the literature the study of Coleman and Vassilicos \cite{coleman2009unified} shows that ZAPS are found for several values of the Okubo-Weiss parameter \cite{okubo1970horizontal}, which gauges the importance of strain over vorticity. More precisely, Coleman and Vassilicos suggest that ZAPS cannot be uniquely associated with regions of high strain and low vorticity. On the other hand,
the study of Bragg et al.~\cite{bragg2015mechanisms} advances that \textit{``regions where the fluid acceleration is low
($\mathbf{a} \to$ 0) are associated with regions where the coarse-grained
strain exceeds the coarse-grained rotation."}}

Moreover, the average cluster size normalized by the average Vorono\"{i} cell size ($\langle V_C/\langle V\rangle \rangle$) shows that clusters for ZAPS and WZERO in Fig.~\ref{fig:CLUM-MEAN} are independent, or at least weakly dependent, on $Re_\lambda$. \textcolor{black}{On the contrary, $\langle V_C/\langle V\rangle \rangle$ for STPS increases with $Re_\lambda$}. To relate these cluster sizes with actual turbulent length scales, we plot in Fig.~\ref{fig:CLUM-LINT} the mean linear size of clusters for all nulls, $\langle V_C\rangle^{1/3}$ against the Kolmogorov and integral lengthscales, respectively. \textcolor{black}{It is worth mentioning that to the authors' best knowledge, this is the first time the clusters of null Lagrangian acceleration and vorticity are characterized via 3D Vorono\"{i} tessellations. For STPS, we see that the cluster average size is two orders of magnitude larger than the Kolmogorov length-scale, somewhat larger than the Taylor length-scale, and grows slowly with $Re_\lambda$. This behavior is to be expected, as the average distance between STPS is related to the Taylor length-scale \cite{goto2009dissipation}, and for HIT $\lambda/\eta\approx 2\sqrt{Re_\lambda}$ \cite{corrsin1963turbulence}. For WZERO, we find that the clusters of null vorticity are consistently smaller than for the other two fields, and they seem to be on the order of 20$\eta$.} For ZAPS we retrieve $\langle V_C\rangle^{1/3}/\eta \approx 30-50$ in lieu of $\langle V_C\rangle^{1/3}/\eta \approx 10-20$ in \cite{Obligado2014}. This mild discrepancy is due to our definition of a cluster: at least two cells ($N_{PC}\geq2$) below the threshold $\mathcal{V}_\textrm{th}$ and that share a face (resp.~edge in 2D) are required to define a cluster. Using this definition yields an average cluster size 2 to 4 times larger than when a condition $N_{PC}\geq1$ is used. Hence, it is thus unsurprising that for ZAPS we obtain a slightly larger value of $\langle V_C\rangle^{1/3}/\eta$ than in other studies.  

\textcolor{black}{In the literature, the origin of these clusters of ZAPS has been suggested to be related to vortical structures. For instance, by means of a 2D numerical study, Faber and Vassilicos \cite{faber2010acceleration} argue that vortical structures centered around ZAPS scale with some lengthscale larger than $\eta$, in fact, a length scale between $\eta$ and $\mathcal{L}$.  If we apply this logic to our observations and assume that clusters of ZAPS scale similarly to those vortical structures identified by Faber and Vassilicos, we can argue that the size of such structures should be close to $0.1\mathcal{L}$}. In addition, the behavior of this quantity is also in agreement with the study of Sumbekova et al.~\cite{sumbekova2017preferential} for inertial particles, which suggests that the average cluster size is an increasing function of the Reynolds number and a fraction of the integral scale.

In the following we continue our analysis characterizing the different shapes of the PDFs of the Vorono\"i cell volumes. As it will be detailed below, not only the global parameters present important differences for STPS, ZAPS, and WZERO, but each set of nulls also has different PDFs and clusters with very different geometrical properties. 

\subsection{Probability density functions of velocity nulls}
\label{sc:pdf}

\textcolor{black}{In this section we consider the PDFs of the Vorono\"i cell volumes and of the volumes of clusters of the velocity nulls in the turbulent field, normalized in all cases by the mean cell volume. To compute each PDF of cluster volumes we used the same volume threshold ($\mathcal{V}_{th}$) for all datasets, and we followed the cluster algorithm of Monchaux et al. \cite{Monchaux2010} to detect clusters, i.e., we took all the neighbouring cells for which $\mathcal{V}<\mathcal{V}_{th}$. The same procedure was used in the following sections to compute PDFs for the volumes of clusters of other vector field nulls.}

\begin{figure}	
	\begin{center}
		
		\begin{subfigure}[t]{0.48\textwidth}
			\begin{overpic}[scale=0.55, trim=0cm 1.5cm 0cm 0cm]{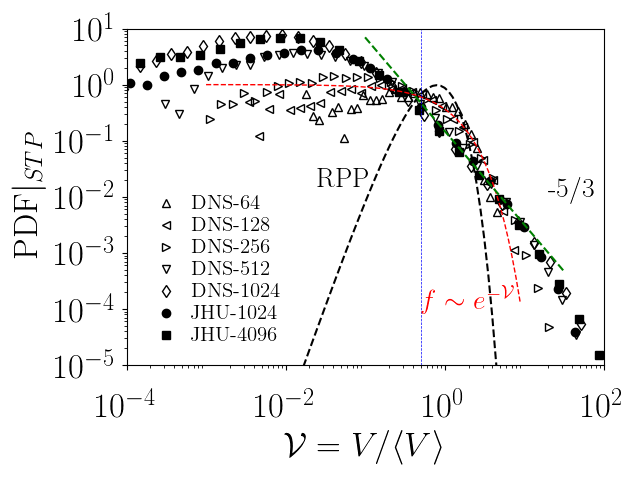}
            \put (85,52) {\huge a)}
            \end{overpic}
			\caption{\label{fig:VOROPDF-STP}}
		\end{subfigure}
		~
		\begin{subfigure}[t]{0.48\textwidth}
			\begin{overpic}[scale=0.55, trim=0cm 1.5cm 0cm 0cm]{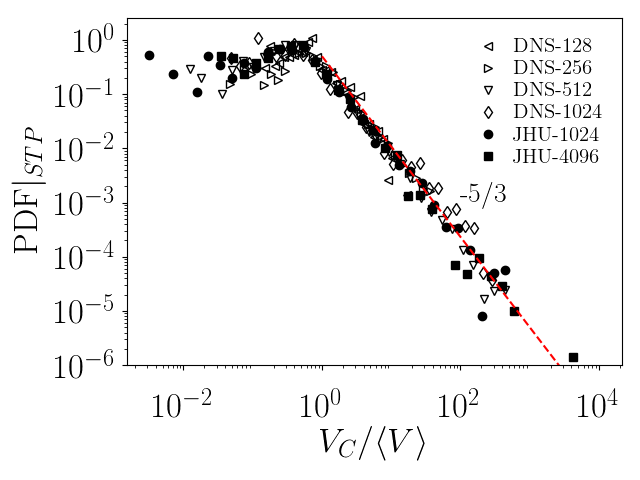}
            \put (25,15) {\huge b)}
            \end{overpic}
			\caption{\label{fig:CLUPDF-STP}}
		\end{subfigure}
		
	\end{center}
	\caption{Stagnation points (STPS) Vorono\"{i} tessellation analysis. a) PDFs of the normalized Vorono\"{i} cell volumes $\mathcal{V}=V/\langle V \rangle$. As a reference, the PDF of an RPP is indicated by the dashed black line, an exponential by a red dashed line, and a $-5/3$ power law by the green dashed line. The vertical dotted line indicates the first crossing of the PDFs with the RPP. Note the similar behavior of the ``DNS" and ``JHU" simulations at the largest resolutions. b) PDFs of the clusters volumes normalized by the average volume ($V_C/\langle V\rangle$), following Monchaux et al.~methodology \cite{Monchaux2010}. A power law is indicated as a reference.}
\end{figure}

The Vorono\"{i} cell volume PDF for velocity nulls exhibits an increasingly wider power-law behavior with an exponent close to $-5/3$ (see Fig.~\ref{fig:VOROPDF-STP}) at increasing values of $Re_\lambda$. \textcolor{black}{This is an expected result: the larger the Reynolds number, the wider the range of scales induced (i.e., the existence of stagnation points separated by different length scales can be just the result of these points being the zeros of a multiscale and approximately self-similar flow). Mora and Obligado \cite{mora2020estimating} also reported the widening power-law behavior with of $Re_\lambda$ to the right of the peak of the PDF in laboratory experiments downstream of an active grid using 1D Vorono\"{i} tessellations. As mentioned before, mathematically this power-law can be a consequence of the power-law behavior of the velocity autocorrelation function. Indeed, Smith et al.~\cite{smith2008fluctuations} found that successive zero crossings (nulls in 1D) of a Gaussian process have power-law behavior with exponent $-2+\beta$ if its spatial autocorrelation function is of the form $\rho(r)=1-{\cal O}(r^{2\beta})$ for $r\ll1$ (see also \cite{orey1970gaussian, davila2003richardson}). The figure also suggests that the crossing at the left of the peak of the PDFs with the PDF of a synthetic random Poisson process (RPP) is somewhat the same for all datasets (as a reference, the crossing is indicated by the vertical blue line in the figure). Thus, one could expect this crossing to be related to some specific length-scale (e.g., to the Taylor length-scale $\lambda$, taking into account that for these nulls $\langle V\rangle^{1/3} = B\lambda$ with $B$ some constant \cite{goto2009dissipation}). However, we did not find conclusive evidence that $\lambda$ is the length scale associated to such crossing. Finally, our volume PDFs also hint that the degree of clustering for STPS scales with $Re_\lambda$, in agreement with the results in the previous section.}

The PDF of the cluster volumes of velocity nulls exhibits an even clearer power-law (see Fig.~\ref{fig:CLUPDF-STP}), also with an exponent close to $-5/3$. The power-law widens over several decades as $Re_\lambda$ increases. However, some previous studies have proposed that this behavior may be trivial or spurious. For instance, Uhlmann and collaborators \cite{uhlmann2014sedimentation, uhlmann2017clustering} have shown that the cluster detection algorithm applied to synthetic random (RPP) data can also yield power-laws. The latter prompts the question of how to differentiate random structures from turbulence driven ones. Mora et al. \cite{MoraIMJF2019} have addressed this problem using a PDF mixture model \cite{fruhwirth2006finite} (see also Sec.~\ref{sec:mixture}). After analyzing the histograms of the number of points inside a cluster ($N_{PC}$), they suggested that this power-law behavior in turbulent flows follows from the functional dependence of these histograms. Mora et al.~further argue that if the probability $P$ of finding a cluster with $N_{PC}$ points goes as $P(N_{PC}) \sim N_{PC}^{\gamma}$, the respective cluster volumes PDF will have a power-law with an exponent close to $\gamma$. Although in 3D the cluster volumes in a RPP may also exhibit such behavior for certain values of the  $\mathcal{V}_\textrm{th}$ threshold, Mora et al.~\cite{MoraIMJF2019} found this behavior is of much wider extent for turbulence-driven clusters. \textcolor{black}{In other words, the collapse seen in the cluster PDFs is a result of the normalization by $\langle V \rangle$, but the extent of the power-law is due to the intensity of the background turbulence, i.e., of the value of $Re_\lambda$, consistent with the argument advanced by Uhlmann \cite{chouippe2019influence}.} 

\subsection{Probability density functions of zero acceleration points and vorticity nulls\label{sec:PDFwzero}}

\begin{figure}	
	\begin{center}
		
		\begin{subfigure}[t]{0.48\textwidth}
		\begin{overpic}[scale=0.55, trim=0cm 1.5cm 0cm 0cm]{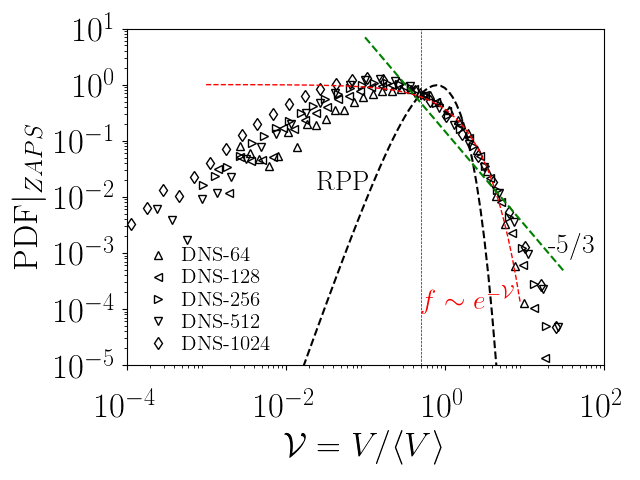}
        \put (85,52) {\huge a)}
        \end{overpic}
    	\caption{\label{fig:VOROPDF-ZAPS}}
		\end{subfigure}
		~
		\begin{subfigure}[t]{0.48\textwidth}
		\begin{overpic}[scale=0.55, trim=0cm 1.5cm 0cm 0cm]{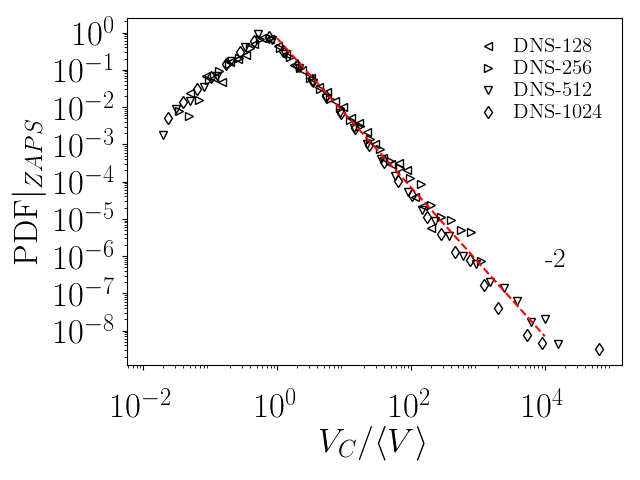}
        \put (25,15) {\huge b)}
        \end{overpic}
		\caption{\label{fig:CLUPDF-ZAPS}}
		\end{subfigure}
		
	\end{center}
	\caption{Zero acceleration points (ZAPS)  Vorono\"{i} tessellation analysis, for the ``DNS" datasets. a) PDFs of the Vorono\"{i} cells normalized volumes $\mathcal{V}=V/\langle V \rangle$. b) PDF of the cluster volumes normalized by the average volume ($V_C/\langle V\rangle$). Power laws and PDFs of exponential and RPP processes are shown as references.}
\end{figure}

Contrary to the STPS, the zero Lagrangian acceleration points PDFs of Vorono\"{i} cell volumes do not exhibit a power-law behavior, and interestingly, when $\mathcal{V}={\cal O}(1)$ they display an almost exponential decay (see Fig.~\ref{fig:VOROPDF-ZAPS}). \textcolor{black}{Once again, the left crossing of the PDFs with the RPP seems to be the same for all datasets}. \textcolor{black}{But as reported in the previous section, the respective ZAPS clusters PDFs exhibit a strong power-law, only this time with an exponent close to $-2$ (see Fig.~\ref{fig:CLUPDF-ZAPS}). The collapse seen in the data is due to the normalization by the average cell volume. Conversely, the extent of this power-law increases with $Re_\lambda$. The behavior of the PDFs of cluster volumes is thus similar to the one observed for the cluster volumes of STPS, but with a different power law.} Obligado et al.~\cite{Obligado2014} report a similar algebraic exponent for ZAPS, and argue it is a signature of the ZAPS clusters' fractal nature. Note that for ZAPS, and for WZERO next, we only report data from the ``DNS" datasets as a result of data availability.

\begin{figure}
	\begin{center}
		
		\begin{subfigure}[t]{0.48\textwidth}
		\begin{overpic}[scale=0.55, trim=0cm 1.5cm 0cm 0cm]{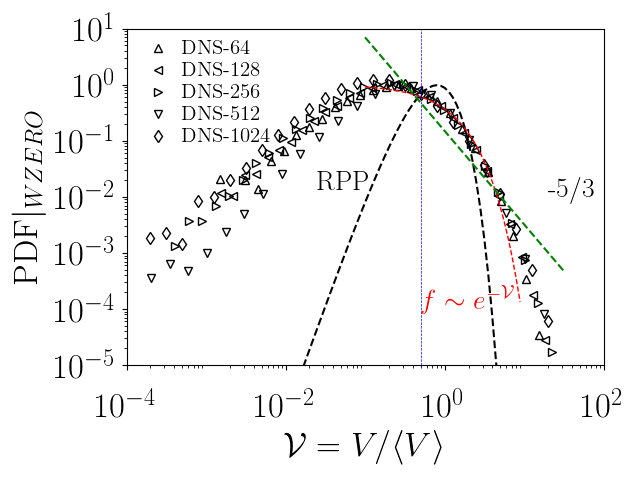}
        \put (85,52) {\huge a)}
		\end{overpic}
    	\caption{\label{fig:VOROPDF-WZERO}}
		\end{subfigure}
		~
		\begin{subfigure}[t]{0.48\textwidth}
		\begin{overpic}[scale=0.55, trim=0cm 1.5cm 0cm 0cm]{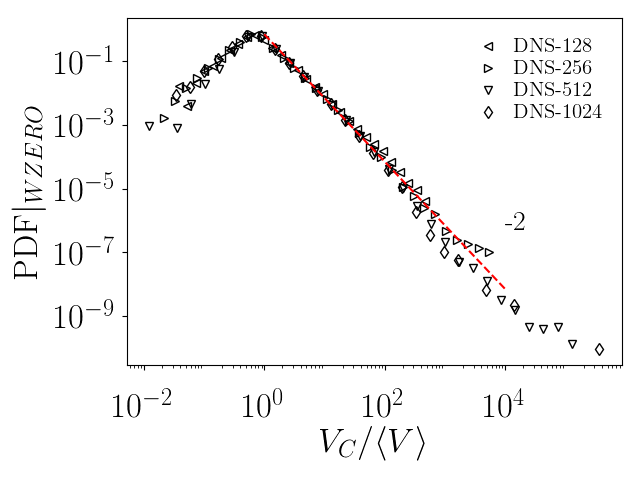}
        \put (25,15) {\huge b)}
		\end{overpic}
		\caption{\label{fig:CLUPDF-WZERO}}
		\end{subfigure}
		
	\end{center}
	\caption{Vorticity null (WZERO) points Vorono\"{i} tessellation analysis. a) PDF of Vorono\"{i} cell volumes $\mathcal{V}=V/\langle V \rangle$. b) PDF of the clusters volumes normalized by the average volume ($V_C/\langle V\rangle$) as in Monchaux et al.~algorithm \cite{Monchaux2010}.}
\end{figure}

\begin{figure}
	\begin{center}
		
		\begin{subfigure}[t]{0.48\textwidth}
		\begin{overpic}[scale=0.55, trim=0cm 1.5cm 0cm 0cm]{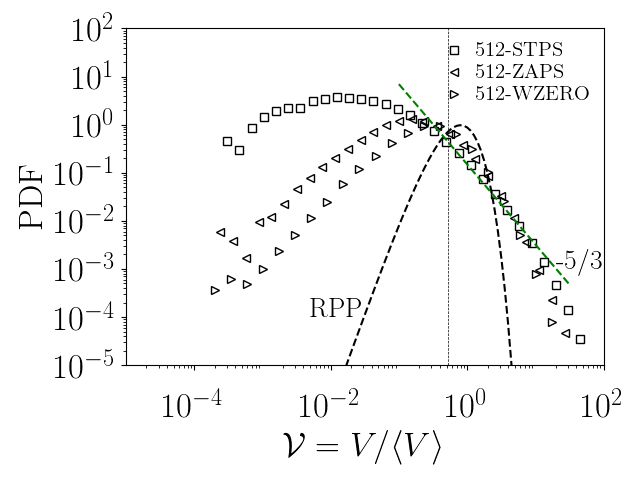}
        \put (25,52) {\huge a)}
		\end{overpic}
		\caption{ \label{fig:VOROPDF-FULL}}
		\end{subfigure}
		~
		\begin{subfigure}[t]{0.48\textwidth}
		\begin{overpic}[scale=0.55, trim=0cm 1.5cm 0cm 0cm]{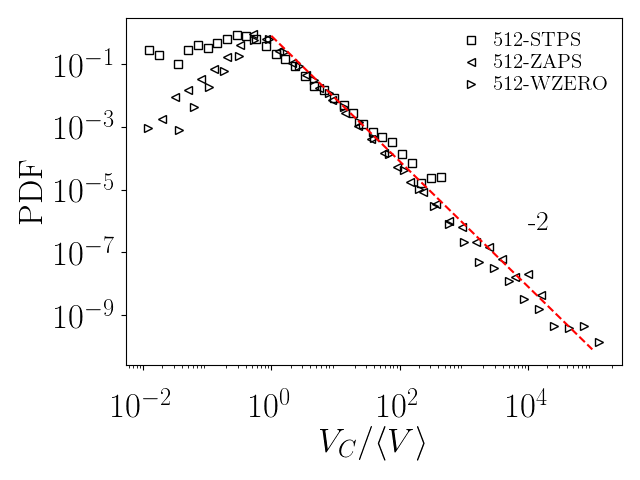}
        \put (25,15) {\huge b)}
		\end{overpic}
		\caption{\label{fig:CLUPDF-FULL}}
		\end{subfigure}
	
	\end{center}
	\caption{Analysis of Vorono\"{i} tessellation for STPS, ZAPS, and WZERO in the DNS-512 simulation. a) PDF of the Vorono\"{i} cells volumes $\mathcal{V}=V/\langle V \rangle$. b) PDF of the clusters volumes normalized by the average volume ($V_C/\langle V\rangle$) using Monchaux et al. algorithm \cite{Monchaux2010}. Power laws and the PDF of a RPP are indicated as references.}
\end{figure}

We now consider the PDFs of Vorono\"{i} cell volumes and of cluster volumes for WZERO. The vorticity nulls Vorono\"{i} cell PDFs display (see Fig.~\ref{fig:VOROPDF-WZERO}) a similar behavior as the one found for the ZAPS Vorono\"{i} cell volume PDF. \textcolor{black}{And qualitatively, the same similarities are observed in the PDFs of the cluster volumes for WZERO.} However, the PDFs of cluster volumes of vorticity nulls have a slightly broader power-law behavior than the respective ZAPS PDFs (see Fig.~\ref{fig:CLUPDF-WZERO}). \textcolor{black}{To confirm these subtle differences a comparison between the three null fields (see Figs.~\ref{fig:VOROPDF-FULL} and \ref{fig:CLUPDF-FULL}) was conducted for the three datasets of nulls in the DNS-512 case (see table \ref{tab:dnspara}). The comparison reveals that: (1) The left crossing between the nulls Vorono\"{i} cell PDFs and the RPP is somewhat similar for all fields (although the position of the peaks are vastly different), (2) the nulls cluster PDFs collapse to some extent, and (3) the extent of the power-law is ordered the different vector nulls as STPS $<$ ZAPS $<$ WZERO (note that for clarity only a $-2$ power law is shown for the PDFs of cluster volumes). In the next section we study the origin of this power-law.}

\subsection{Power laws in the probability density functions of the cluster volumes\label{sec:mixture}}

We now examine the PDFs of ZAPS and WZERO clusters volumes using the approach of Mora et al.~\cite{MoraIMJF2019}. These authors claim that clusters PDFs, obtained by the clustering detection algorithm in Monchaux et al.~\cite{Monchaux2010} (see also Sec.
~\ref{sc:nall}) can be analytically described by a mixture PDF model \cite{fruhwirth2006finite}. PDF mixture models are based on PDFs linear superpositions: Invididual PDFs $f_i$ are multiplied by weights $\alpha_i$, i.e., $f_{mix}=\sum_i^{N_P} \alpha_i f_i$, where $N_P$ is the number of PDFs to combine. For instance, $N_P$ can be associated with the number of points (\textcolor{black}{resp. particles}) in the clusters, and the PDFs of clusters of two, three, four, and up to $N_P$ points can be combined to construct a PDF which represents the statistics of $V_C/\langle V\rangle$. These $N_P$-points cluster PDFs are computed via convolutions (i.e., assuming statistical independence and strong-mixing conditions \cite{ibragimov1975note, bradley1981central}) using a limited 3D random Poisson distribution as base function. For more details, see \cite{MoraIMJF2019}.

\begin{figure}
	\begin{center}	
	
		\begin{subfigure}[t]{0.48\textwidth}
		\begin{overpic}[scale=0.55, trim=0cm 1.5cm 0cm 0cm]{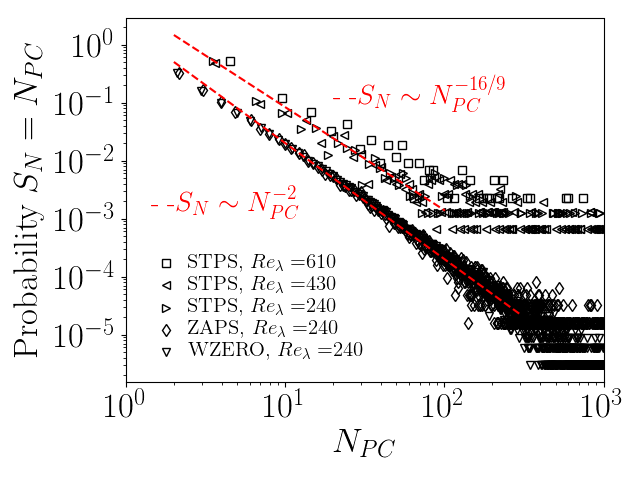}
        \put (85,52) {\huge a)}
		\end{overpic}
        \caption{ \label{fig:HISTO}}
		\end{subfigure}
		~
		\begin{subfigure}[t]{0.48\textwidth}
		\begin{overpic}[scale=0.55, trim=0cm 1.5cm 0cm 0cm]{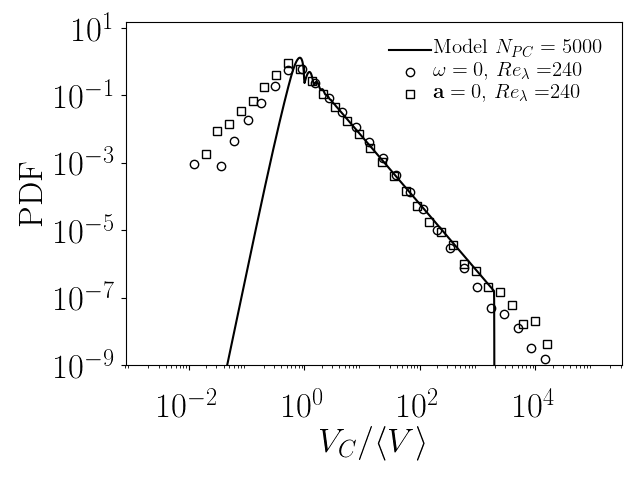}
        \put (25,15) {\huge b)}
		\end{overpic}
		\caption{ \label{fig:MODEL-PDF}}
		\end{subfigure}
	
	\end{center}
	\caption{a) Probability histogram $S_N$ of the number of clusters with $N_{PC}$ points inside the clusters, for STPS, ZAPS and, WZERO, at different values of $Re_\lambda$. Power laws are shown as references. b) PDF of clusters volumes of ZAPS and WZERO including the PDF from the model proposed by Mora et al.~\cite{MoraIMJF2019} with $\mathcal{V}_\textrm{th}=0.5$ and $N_{PC}$ up to 5000.}
\end{figure}

Mora et al.~\cite{MoraIMJF2019} suggest computing the weights $\alpha_i$ as $\alpha_{N_{PC}}=$ number of clusters with $N_{PC}$ points divided by the total number of clusters. Thus, we estimated these weights by computing histograms ($S_N$) of the number of clusters conditioned on the number of null points (\textcolor{black}{resp.~particles}) in a cluster $N_{PC}$ (see Fig.~\ref{fig:HISTO}). These histograms have a power-law behavior with an exponent close to $-16/9$ or to $-2$ (maybe slightly dependent on the field nulls considered), and similar to the exponent proposed by Yoshimoto and Goto \cite{yoshimoto2007self} for inertial particles. The observation of this scaling cannot be overlooked, as it gives credence to some aspects proposed by the sweep-stick mechanism: particle clustering is a multi-scale process that resembles the clustering of acceleration nulls. Indeed, it is remarkable that previous studies with inertial particles \cite{goto2006self} found a compatible self-similar structure to that followed by these pairs, triplets, quartets, and so on of vector nulls.

\textcolor{black}{We thus fed the mixture PDF model with $\alpha_{PC}\sim N_{PC}^{-2}$ (i.e., the probability of having clusters of $N_{PC}$ points follows a power-law), and computed the PDF summing up to $N_{PC}=5000$ with $\mathrm{max}(V_C/\langle V\rangle)\approx 5000\mathcal{V}_\textrm{th}\approx 2500$ for $\mathcal{V}_\textrm{th} = 0.5$. For clarity we only compare the result of this mixture PDF model with the cluster volume PDFs for ZAPS and WZERO, which have a wider power-law extent (c.f.~Fig.~\ref{fig:CLUPDF-FULL}), and which also show a more compatible behavior with the assumption that $\alpha_{PC}\sim N_{PC}^{-2}$. The result is shown in Fig.~\ref{fig:MODEL-PDF}.} Indeed we find a good agreement, recovering the power-law behavior of the PDFs of ZAPS and WZEROS for several decade (also shown in Fig.~\ref{fig:MODEL-PDF}).

This result is remarkable considering that the PDF mixture model uses convolutions of limited RPP distributions, each of them with no correlation at any scale \cite{ferenc2007size}. However, their superposition can have correlations given by the coefficients in the expansion. Therefore, the broader power-law behavior seen in the clusters PDFs (Figs.~\ref{fig:CLUPDF-STP}, \ref{fig:CLUPDF-ZAPS}, and \ref{fig:CLUPDF-WZERO}) can only have a turbulent origin. Indeed, its origin resides in the power law scaling of the $\alpha_i$ weights. The resulting behavior, although it may depend on the threshold $\mathcal{V}_\textrm{th}$ used, is therefore not a spurious artifact of the 3D Vorono\"{i} tessellations (see also Fig.~13 in \cite{chouippe2019influence}); turbulence increases the probability --through the weights $\alpha_i$-- of having very large structures as those found in STPS, ZAPS, or WZERO. 


\section{Inertial particles in HIT}\label{sec:part}

As previously mentioned, Coleman and Vassilicos \cite{coleman2009unified} have related the geometry of ZAPS to inertial particle clustering (also known as preferential concentration). To examine this phenomenon, we tracked point inertial particles in DNS-512 with Stokes numbers $St$ equal to $0.5$, 3, and 6 respectively. Each dataset of particles (for each value of $St$) comprised 20 snapshots containing instantaneous 3D positions of 10$^6$ inertial particles.

\begin{figure}
	\begin{center}
		
		\begin{subfigure}[t]{0.48\textwidth}
		\begin{overpic}[scale=0.55, trim=0cm 1.5cm 0cm 0cm]{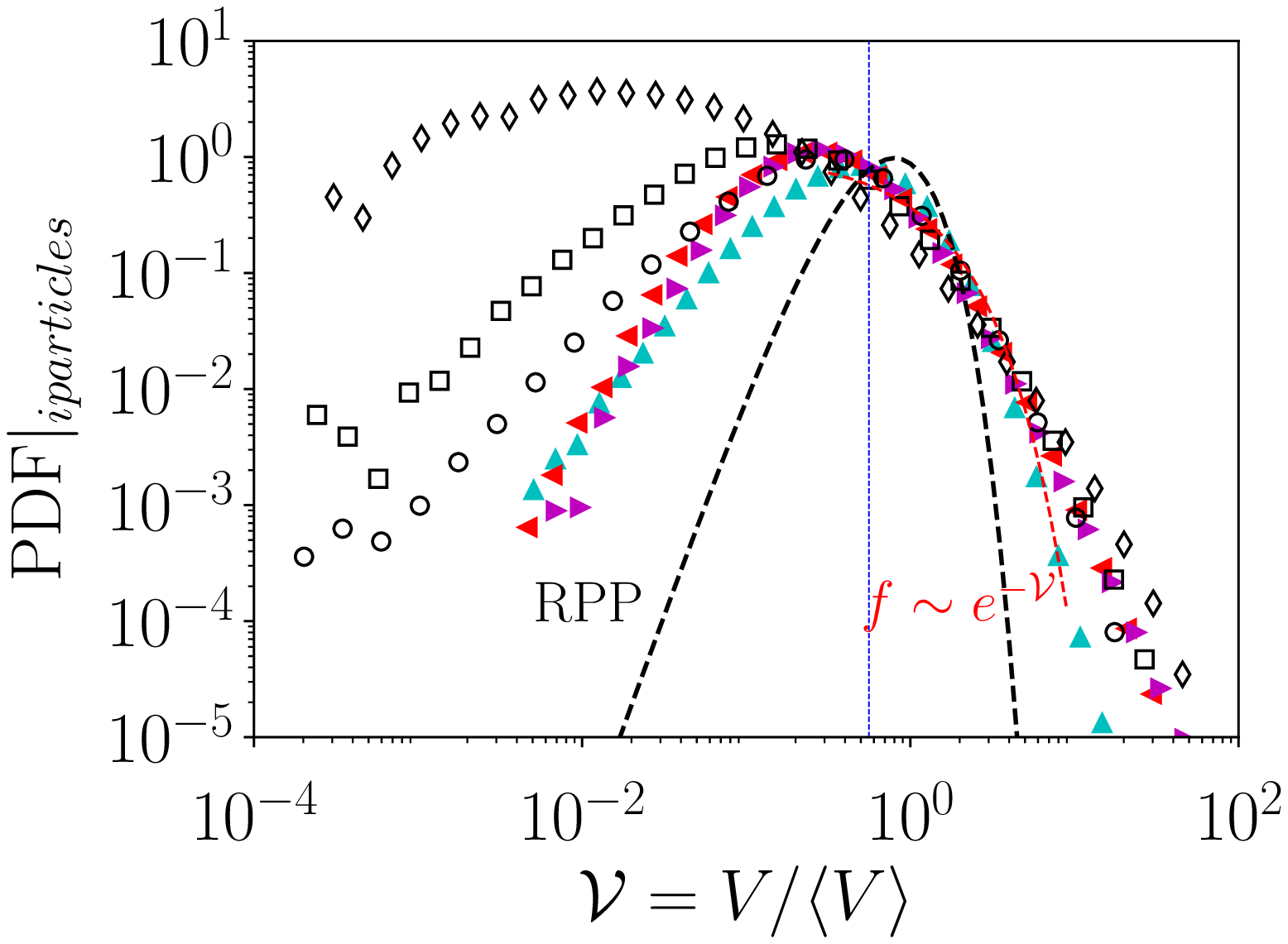}
        \put (85,52) {\huge a)}
		\end{overpic}
		\caption{ \label{fig:VOROPDF-IP}}
		\end{subfigure}
		~
		\begin{subfigure}[t]{0.48\textwidth}
		\begin{overpic}[scale=0.55, trim=0cm 1.5cm 0cm 0cm]{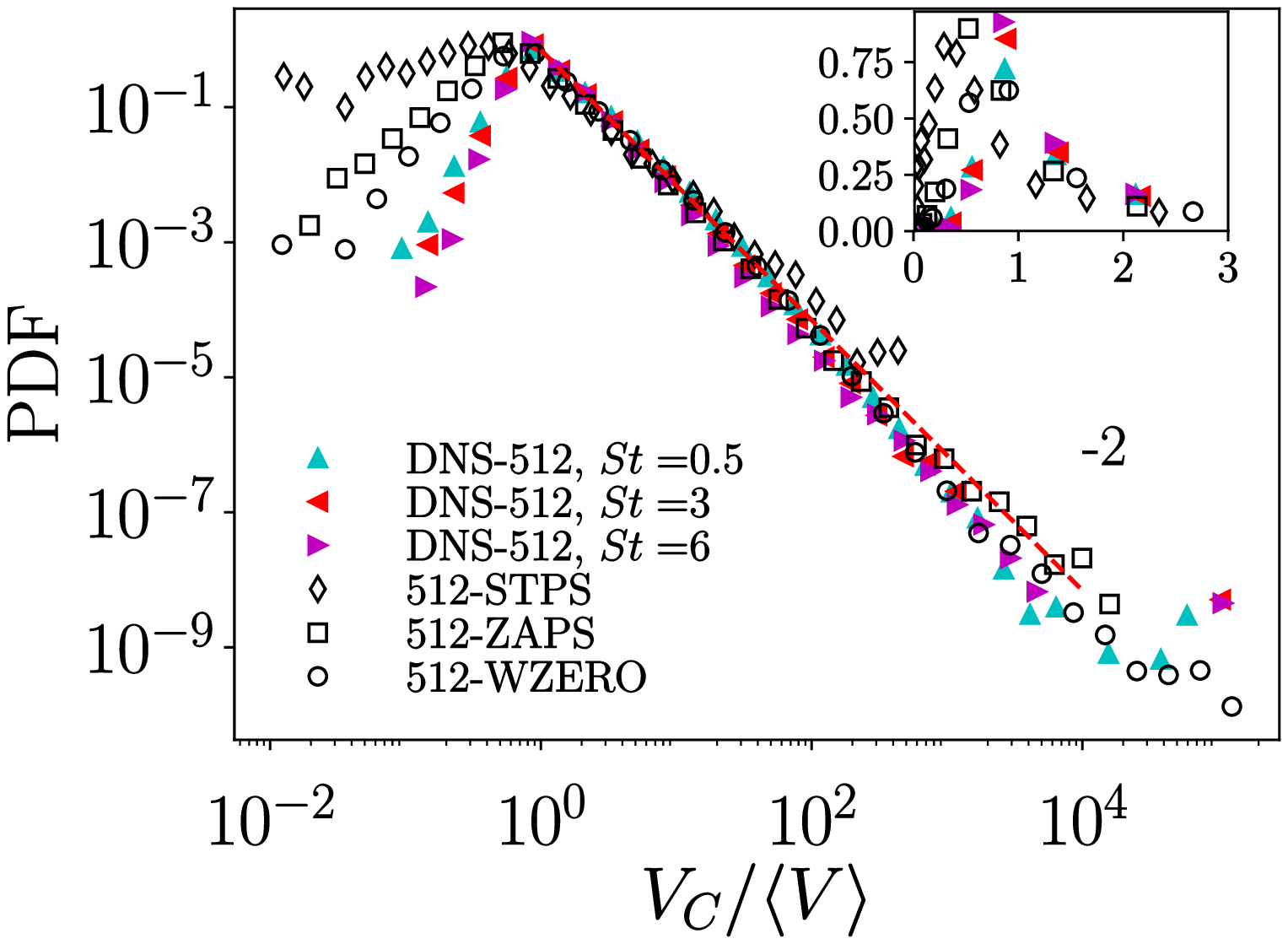}
        \put (84,54) {\huge b)}
		\end{overpic}
		\caption{\label{fig:CLUPDF-IP}}
		\end{subfigure}
		
		\begin{subfigure}[t]{0.48\textwidth}
		\begin{overpic}[scale=0.55, trim=0cm 1.5cm 0cm 0cm]{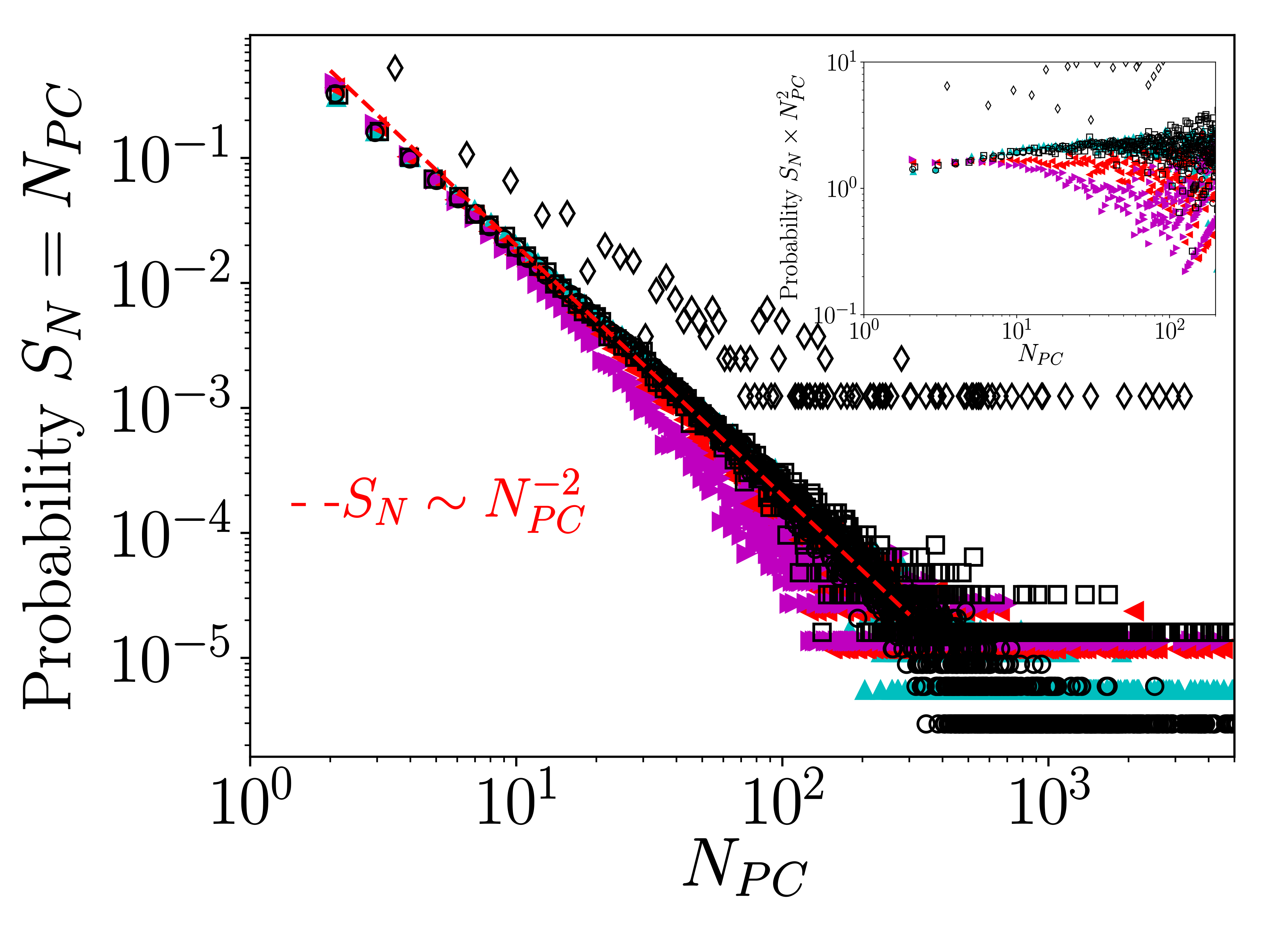}
        \put (25,12) {\huge c)}
		\end{overpic}
		\caption{\label{fig:HISTO-PAR}}
		\end{subfigure}
        ~
		\begin{subfigure}[t]{0.48\textwidth}
		\begin{overpic}[scale=0.55, trim=0cm 1.5cm 0cm 0cm]{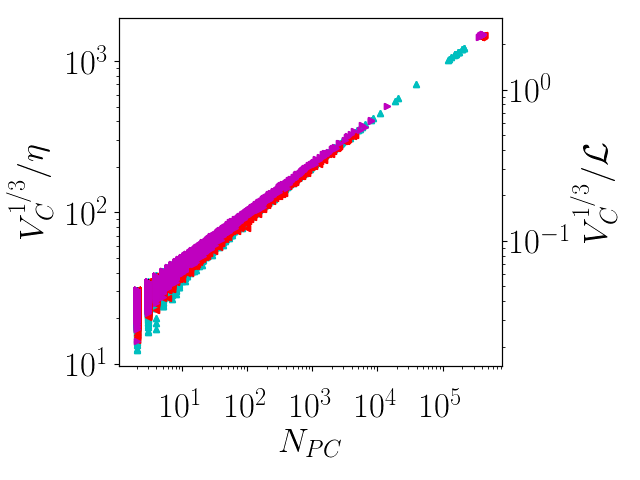}
        \put (25,52) {\huge d)}
		\end{overpic}
		\caption{\label{fig:VOROPDF-PVS-IP}}
		\end{subfigure}
		
	\end{center}
	\caption{Vorono\"{i} tessellation analysis of inertial particles' instantaneous positions. a) PDF of the normalized Vorono\"{i} cells volumes $\mathcal{V}=V/\langle V \rangle$. Markers follow the legend in panel b). b) PDF of the clusters volumes normalized by the average volume ($V_C/\langle V\rangle$) using Monchaux et al.~criteria. The inset shows a detail of the peak of the PDFs in linear scale. Open markers refer to STPS, ZAPS and WZERO in DNS-512, while closed markers are for particles with different Stokes numbers. c) Histogram of number of clusters conditioned on the number of points in the cluster, for STPS, ZAPS, WZERO, and particles with different $St$. \textcolor{black}{The inset shows the compensated histogram by an algebraic exponent of 2, i.e., the PDFs in the main figure multiplied by $N_{PC}^2$}. d) Linear cluster size (normalized by the Komogorov and integral lengths) against the number of particles inside the cluster. In these panels, RPP distributions and power laws are shown as references.}
\end{figure}

\begin{figure}	
	\begin{center}
	
		\begin{subfigure}[t]{0.48\textwidth}
		\begin{overpic}[scale=0.55,trim=0cm 1.5cm 0cm 0cm]{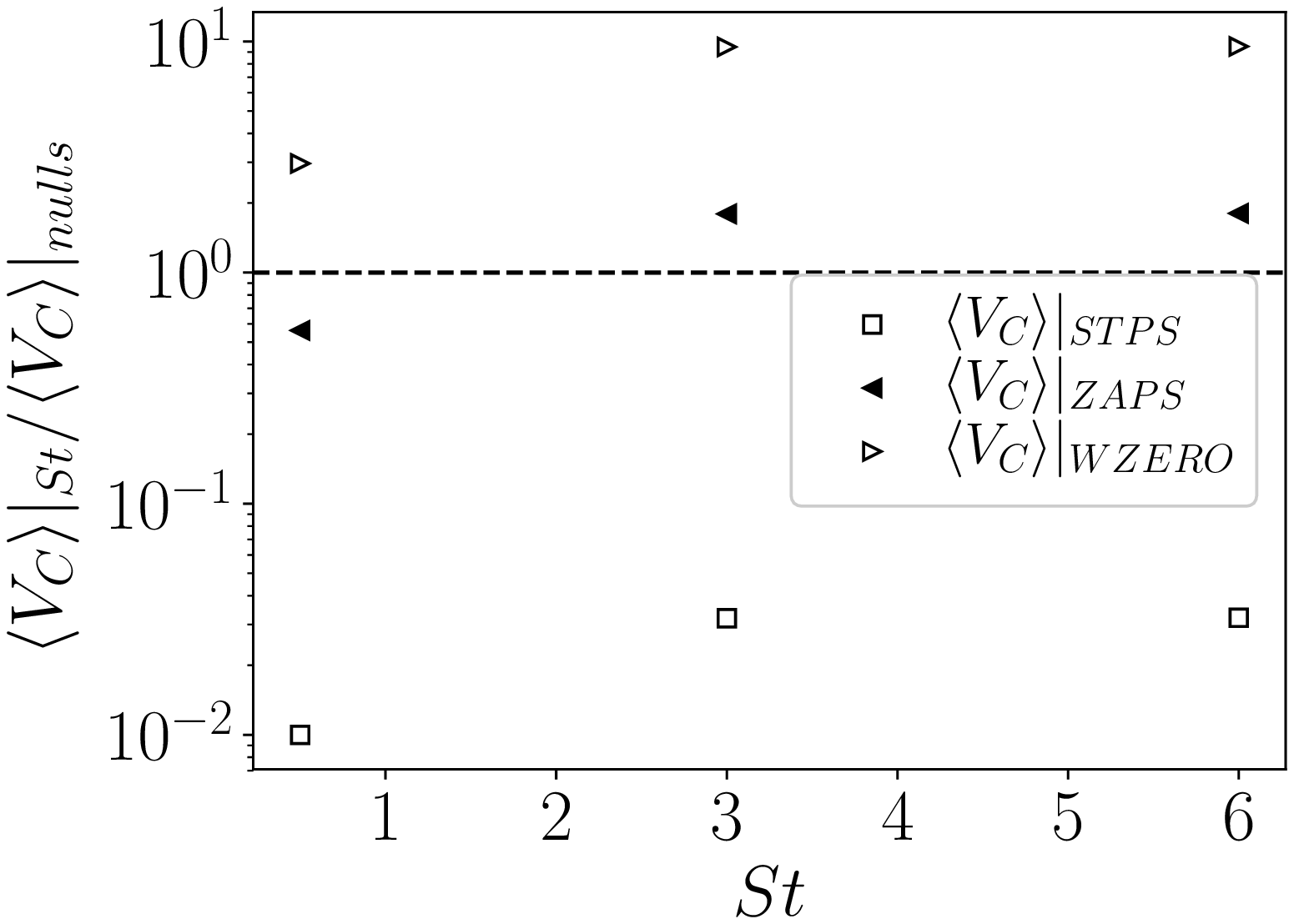}
        \put (23,20) {\huge a)}
		\end{overpic}
		\caption{\label{fig:CLUM-MEAN-ST}}
		\end{subfigure}
        ~
		\begin{subfigure}[t]{0.48\textwidth}
		\begin{overpic}[scale=0.55,trim=0cm 1.5cm 0cm 0cm]{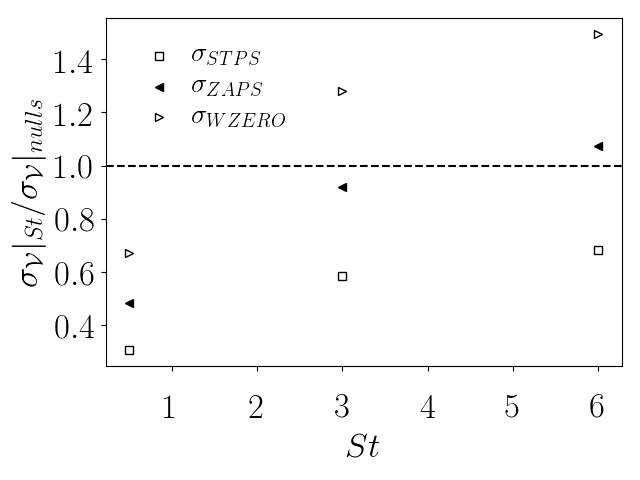}
        \put (88,14) {\huge b)}
		\end{overpic}
		\caption{\label{fig:STD-COMP}}
		\end{subfigure}

	\end{center}
	\caption{Particle clusters global parameters. a) Inertial particles average cluster size over the  different average cluster sizes coming from the nulls of fields of the DNS-512 dataset. b) Ratio of the standard deviation of the inertial particle cluster volumes over the standard deviation for the different nulls fields, also for the DNS-512 dataset. }
\end{figure}

We found evidence of clustering of particles for all Stokes numbers considered (see Fig.~\ref{fig:VOROPDF-IP}, which also shows a comparison with the PDFs of cell volumes of STPS, ZAPS and WZERO). Interestingly, the particles Vorono\"i volumes PDF have better agreement with the WZERO PDF for $\mathcal{V}\ll1$, and on the contrary, they exhibit better agreement with the ZAPS PDF for $\mathcal{V}\gtrsim 1$ \textcolor{black}{(albeit not as good for the particles with $St = 0.5$, as evidenced by a slightly faster drop of the PDF of these particles for $\mathcal{V}\gtrsim 5$). But given the overall similarities of the PDFs for $\mathcal{V}\gtrsim 1$}, it is unsurprising that the inertial particles degree of clustering shows an overall better agreement with the ZAPS field (as later shown in Fig.~\ref{fig:STD-COMP}), as larger cells, also known as \textit{voids}, are the main contributors to the standard deviation of the cell volumes $\sigma_\mathcal{V}$ \cite{sumbekova2017preferential}. However, all these observations also suggest that the \textit{sweep-stick} mechanism may only be an approximate representation of the phenomenon underlying the physics at all scales, as discussed by Bragg et al.~\cite{bragg2015mechanisms}. 

The behavior for $\mathcal{V}\ll1$ has not been observed by the Obligado et al.~study \cite{Obligado2014}, which used experimental data for inertial particles (using 2D high-speed imaging) and patches of low Lagrangian acceleration obtained from DNS. These patches were also averaged into 2D planes with several pixels thickness (to mimic the finite thickness of a laser sheet in the experiments). Therefore, both experiments (due to finite spatial resolution) and DNS (due to the averaging) did not resolve the range $\mathcal{V} \ll1$ corresponding to very small cells. For larger cells, a statistical agreement between particles clustering and ZAPS was reported, in agreement with the present results. \textcolor{black}{As will be explained below, we believe that for physical reasons the $\mathcal{V} \ll1$ range is less relevant, at least for the simulations we consider and the approximations we made in the dynamics of the inertial particles.}

Interestingly, we also retrieve a power law (see Fig.~\ref{fig:CLUPDF-IP}) in the PDF of cluster volumes of particles, analogous to the one found for ZAPS and WZERO. The cluster volume PDFs peak at $V_C/\langle V\rangle\approx1 $ can be explained from the histograms of number of clusters conditioned on the number of particles inside a cluster $N_{PC}$ (Fig.~\ref{fig:HISTO-PAR}); clusters with $N_{PC}=2$ have a larger probability, and therefore, $V_C/\langle V\rangle\approx2\mathcal{V}_\textrm{th}\approx1$ is the most likely value. Considering that the normalized cells $\mathcal{V}$  close to the threshold $\mathcal{V}_\textrm{th}$ have the highest probability, our observation is insensitive to the reported linear behavior between $N_{PC}$ and cluster size (see Fig.~\ref{fig:VOROPDF-PVS-IP} and \cite{momenifar2020local}).
\textcolor{black}{Also, note that all the PDFs in Fig.~\ref{fig:CLUPDF-IP} are normalized by the respective mean cell volume. For clarity, the actual mean value and dispersion of the cluster sizes (compared with those of STPS, ZAPS and WZERO) will be discussed later, for which a dependence with $St$ will be more clearly observed. Closer inspection of the histogram in Fig.~\ref{fig:HISTO-PAR} reveals good agreement between the different particles sets and ZAPS or WZEROS for $St=0.5$, $St=3$ and $St=6$ up to $N_{PC}<10$. In other words, for clusters with $N_{PC}>10$ a dependence on the value of $St$ of the particles can also be observed (see the inset in \ref{fig:HISTO-PAR}, which shows the PDFs compensated by $N_{PC}^2$). In particular, beyond this value the largest Stokes numbers start to depart from the smallest one.} This discrepancy agrees with the expectation that at larger Stokes numbers the particles filter out certain flow scales \cite{bec2006acceleration, Angriman_2020}. In particular, for large values of $St$, particles are expected to be less sensitive to fast changing (and small scale) motions in the fluid.

\textcolor{black}{Considering again the PDF of cluster volumes, note also that the probability of having small clusters $V_C/\langle V \rangle\ll1$ (see Fig.~\ref{fig:CLUPDF-IP}) is higher for all the vector nulls than for the particles, a behavior similar to that observed in Fig.~\ref{fig:VOROPDF-IP} for the Vorono\"i cell volumes). Although dispelling why this is the case is interesting, we refrain from analyzing these high concentration regions \mica{as numerical models for inertial particles simulations may reach, there, physical limits}. In these regions, the particle local concentration can indeed be orders of magnitude larger than the surrounding conditions \cite{huck2018role,aliseda2002effect} \mica{(as a cluster of $N_{PC}$ particles such that $V_C/\langle V\rangle <0.1 $ is for instance at least $10 N_{PC}$ times denser than the average seeding concentration $\langle V \rangle^{-1}$)}. Therefore, the assumption of the one-way coupling approximation may become invalid in these regions as the particle clusters should actually strongly modify the surrounding turbulence (though a so called \textit{two-way} coupling) in a real flow \cite{elghobashi1994predicting, balachandar2010turbulent, MoraPRL2019}. \mica{To be more specific, in the present study where $10^6$ particles are simulated in a $(2\pi)^3$ numerical volume, the linear dimension of a cluster such that $V_C/ \langle V \rangle = 0.1$ is $l_c\simeq 0.006$, hence about the size of $\eta$ for the 512$^3$ simulation. Besides, by the definition of the Stokes number, the diameter of the particles relates to $\eta$ by $d_p=(6 \, St^{1/2} \, \eta)/\sqrt{1+2 \Gamma}$, where $\Gamma=\rho_p/\rho_f$ is the particle to fluid density ratio. Therefore, if we were for instance interested in water droplets in air (a problem relevant for clouds), particles with $St\simeq 1$ are such that $d_p \simeq 0.15 \eta$. As a result, a cluster such that $V_C/\langle V \rangle = 0.1$ can realistically contain at most 5 to 6 particles, which are then fully compact (hence at odds with the one-way coupling approximation), while a cluster such that $V_C < 0.01 \langle V\rangle$ should contain at most 1 particle, which then becomes irrelevant. Finally, these small clusters tend to be either clusters with very few particles (which just happened to be sporadically close, but cannot be relevantly considered as coherent clusters~\cite{baker2017coherent}) or too highly seeded (hence out of the physical approximations of numerical models). This justifies, that when it comes to take into consideration physical constraints in real particle laden flows, such small or presumably too highly concentrated clusters, which are artificially accessible in numerical simulations, must not be over-considered.} As a result we focus mostly on the self-similar clusters \mica{($V_C/\langle V \rangle > 1$)} as done, e.g., by Baker et al.~\cite{baker2017coherent, petersen2019experimental}.}

Interestingly, inertial particle clusters volumes are (on average): (1) Smaller than the average volume of STPS clusters, (2) of the same order of ZAPS clusters, and (3) larger than WZERO clusters (see Fig.~\ref{fig:CLUM-MEAN-ST}, which compares the ratios of the mean particles cluster volumes to those of the three nulls). \textcolor{black}{However, the ratios between these volumes are not completely independent of the Stokes number of the particles.} The ratios seem to grow with $St$ until saturating, \textcolor{black}{and for particles with $St = 0.5$ the mean volume of the clusters is somewhere in between those of WZERO and ZAPS, being $\approx 3$ times larger than the mean volume of clusters of WZERO, and $\approx 2$ times smaller than the mean volume of clusters of ZAPS. A qualitatively similar behavior is observed in the standard deviation of the PDFs of $\mathcal{V}$ (the normalized Vorono\"{i} cell volumes) for the particles with different $St$ compared with the standard deviation in $\mathcal{V}$ for STPS, ZAPS, and WZERO (see Fig.~\ref{fig:STD-COMP}). In particular, note that for $St = 3$ and 6, $(\sigma_{\mathcal{V}}|_{St})/(\sigma_{\mathcal{V}}|_{ZAPS}) \approx 1$. Given that the standard deviation of the Vorono\"{i} tessellation quantifies the degree of clustering, whereas the PDFs of clusters depend on the probability of finding clusters of 2, 3, 4, and larger number of particles (as shown in Sec.~\ref{sec:mixture} and in Fig~ \ref{fig:MODEL-PDF}), the statistical similarities between the clustering of ZAPS and of inertial particles are again remarkable. Indeed, both quantities have average cluster volumes of the same order of magnitude, i.e., $(\langle V_C \rangle\vert_{St})/(\langle V_C \rangle \vert_{ZAPS}) \in [0.5-2]$} as already reported, and specially so for the particles with $St>1$. Obligado et al.~\cite{Obligado2014} reported a similar trend from experimental measurements taken via 2D imaging. The ratio of the standard deviations (which, as just mentioned, provides a way to quantify the strength of the clustering \cite{Monchaux2010}) in Fig.~\ref{fig:STD-COMP} is also in agreement with the results of  Obligado et al.~\cite{Obligado2014}. This observation further supports the observation that the degree of clustering of inertial particles for $St\geq1$ has indeed a close resemblance to self-similar clustering of ZAPS.

We therefore find that, while the 3D nature and more detailed spatial resolution of this study shows that no set of nulls perfectly mimics the inertial particles spatial distribution, for $St \geq 1$ the results from~\cite{Obligado2014} remain valid: the spatial segregation of particles is consistent with that of ZAPS (specially for larger Vorono\"{i} volumes), supporting an agreement with the sweep-stick mechanism. On the other hand, very small particles' Vorono\"{i} cells deviate from this behavior, and particles with $St = 0.5$ do not present complete similarities with any of the set of nulls discussed here. It is nevertheless expected that for lower values of the $St$ number, inertial particles will eventually agglomerate within low vorticity regions of the flow.

As a closing comment, note other scaling relations can be inferred or confirmed from these results. Previous studies \cite{monchaux2017settling, MoraIMJF2019} indicate that the linear cluster size of inertial particles $L_C = \langle V_C \rangle\vert_{St}^{1/3}$ is of $\mathcal{O}(\mathcal{L}/10)$. Our results also indicate $L_C/\mathcal{L}\in [0.1-0.15]$, with similar values for linear cluster sizes of ZAPS for $Re_\lambda>200$. In a related observation, Wittmeier and Shrimpton \cite{wittemeier2018explanation} recently reported that when the product between the particle number density and the Kolmogorov length scale is held constant, i.e., $n_p^{1/3}\eta=\eta/\langle V\rangle\vert_{particle}^{1/3}=\textrm{constant}$, some measures used to quantify preferential concentration become independent of the Reynolds number for $Re_\lambda\geq200$. We can test this claim in the following way: first, we assume that the particles follow the \textit{sweep-stick} mechanism, and that the number density of ZAPS and of inertial particles are similar such that we can write $n_p^{1/3}\eta\sim n_{ZAPS}^{1/3}\eta$. Then, using Fig.~\ref{fig:MEANVORO-F} (see also \cite{chen2006turbulent}) we can advance $n_{ZAPS}\sim(\mathcal{L}/\eta)^{3}$ and thus $n_p^{1/3}\eta \sim n_{ZAPS}^{1/3}\eta \sim \mathcal{L}$. Our DNS results supports this proposal, and the degree of clustering of ZAPS appears to saturate for $Re_\lambda \in [250,610]$ , i.e., $(\sigma_\mathcal{V}\vert_{ZAPS})/\sigma_{RPP}\approx 4$.

\section{Concluding remarks}

We have analyzed the velocity, Lagrangian acceleration, and vorticity nulls in datasets coming from high fidelity numerical simulations, in a wide range of Taylor-based Reynolds numbers. Mean values and standard deviations of Vorono\"{i} cells volumes for these fields nulls display scaling dependence with $Re_\lambda$. The number density of the velocity and acceleration nulls roughly follow the scalings proposed by Vassilicos and collaborators \cite{davila2003richardson, goto2004particle, chen2006turbulent}. Vorticity nulls (the densest of all fields) also exhibit a scaling similar to the acceleration nulls, as reported by Moisy and Jimenez\cite{moisy2004geometry}. The velocity nulls are scarce, but they are the most strongly clustered field at increasing $Re_\lambda$, as indicated by the standard deviation of Vorono\"{i} cells volumes. On the contrary, clustering of vorticity and Lagrangian acceleration nulls (again as indicated by the standard deviations) barely changes with $Re_\lambda$, with their normalized cluster size depending weakly on $Re_\lambda$.

Our results confirm the presence of a power-law with an exponent close to $-5/3$ in the Vorono\"{i} volume cell PDF for velocity nulls (or stagnation points) at increasing values of $Re_\lambda$. This behavior is absent for acceleration and vorticity nulls. Moreover, when considering the PDFs of cluster volumes, the PDFs for all null fields show a power-law behavior with an algebraic exponent close to $-5/3$ for velocity nulls, and to $-2$ for Lagrangian acceleration and vorticity nulls. We showed evidence that this behavior is not an artifact of the 3D Vorono\"{i} tessellation, and that the extent of the scaling stems from the underlying dynamics of the turbulent flow.

When considering the clustering (or preferential concentration) of point inertial particles, our results show that for Vorono\"i cells with normalized volume $\mathcal{V}>1$ (i.e., for volumes larger than the mean), the Vorono\"{i} cell PDF of inertial particle clustering better matches the ZAPS Vorono\"{i} cell PDF, \textcolor{black}{specially for the particles considered with $St>1$.} Likewise, the average cluster volume of both inertial particles and ZAPS have the same order of magnitude for these particles. These observations give credence to the observation that on the average, the preferential concentration mimics the topology of the zero acceleration points, as reported elsewhere \cite{coleman2009unified, Obligado2014}. However, for very small particles' Vorono\"{i} cells and for particles with $St=0.3$, deviations from this behavior are observed, indicating that the {\it sweep-stick} mechanism may be only an an approximate representation of a more complex physical process underlying the preferential concentration of particles. Finally, we find evidence that the cluster linear size scales with the integral length scale, $L_C=\mathcal{O}( \mathcal{L}/10)$, in agreement with previous studies by Mora et al.~\cite{MoraIMJF2019}. 

\begin{acknowledgments}
This work was partially supported by the ECOS project A18ST04. D.O.M.~and M.O.~acknowledge the LabEx Tec21 (Investissements d'Avenir - Grant Agreement $\#$ ANR-11-LABX-0030), and the ANR project ANR-15-IDEX-02 for funding this work. P.D.M.~acknowledges support from grants PICT No.~2015-3530 and 2018-4298. This research made use of the SciServer science platform (www.sciserver.org). SciServer is a collaborative research environment for large-scale data-driven science. It is developed at, and administered by, the Institute for Data Intensive Engineering and Science at Johns Hopkins University. SciServer is funded by the National Science Foundation through the Data Infrastructure Building Blocks (DIBBs) program and others, as well as by the Alfred P.~Sloan Foundation and the Gordon and Betty Moore Foundation.
\end{acknowledgments}

\providecommand{\noopsort}[1]{}\providecommand{\singleletter}[1]{#1}%

\end{document}